\documentclass[conference]{IEEEtran}
\IEEEoverridecommandlockouts
\usepackage{cite}
\usepackage{amsmath,amssymb,amsfonts}
\usepackage{algorithmic}
\usepackage{graphicx}
\usepackage{textcomp}
\usepackage{xcolor}
\def\BibTeX{{\rm B\kern-.05em{\sc i\kern-.025em b}\kern-.08em
    T\kern-.1667em\lower.7ex\hbox{E}\kern-.125emX}}

\usepackage[ruled]{algorithm}
\usepackage{algorithmic}
\DeclareMathOperator{\offspring}{\mathit{offspring}}
\DeclareMathOperator{\evaluateFitness}{\mathit{evaluateFitness}}
\DeclareMathOperator{\maxFitness}{\mathit{maxFitness}}
\DeclareMathOperator{\getMaxFitness}{\mathit{getMaxFitness}}
\DeclareMathOperator{\indfitness}{\mathit{ind.fitness}}
\DeclareMathOperator{\generateOffspring}{\mathit{generateOffspring}}
\usepackage{eqparbox}

\floatname{algorithm}{Algorithm}

\usepackage{hyperref}

\begin{document}

\title{A Parallel Novelty Search Metaheuristic Applied to a Wildfire Prediction System\\
  \thanks{This work has been funded by CONICET (Argentinean Council for Scientific and Technological Research), by a postdoctoral scholarship for the first author.}
}

\author{\IEEEauthorblockN{Jan Strappa}
\IEEEauthorblockA{\textit{Consejo Nacional de Investigaciones}\\
\textit{Científicas y T\'ecnicas (CONICET)} \\
Mendoza, Argentina} 
\IEEEauthorblockA{\textit{Facultad Regional Mendoza} \\
\textit{Universidad Tecnológica Nacional}\\
Argentina \\
jstrappa@frm.utn.edu.ar \\
ORCiD ID: 0000-0003-3008-0905 
}
\and 
\IEEEauthorblockN{Paola Caymes-Scutari}
\IEEEauthorblockA{\textit{Consejo Nacional de Investigaciones}\\
\textit{Científicas y T\'ecnicas (CONICET)} \\
Argentina} 
\IEEEauthorblockA{\textit{Facultad Regional Mendoza} \\
\textit{Universidad Tecnológica Nacional}\\
Mendoza, Argentina \\
pcaymesscutari@frm.utn.edu.ar\\
ORCiD ID: 0000-0002-6792-0472
}
\and
\IEEEauthorblockN{Germ\'an Bianchini}
\IEEEauthorblockA{\textit{Facultad Regional Mendoza} \\
\textit{Universidad Tecnológica Nacional}\\
Argentina \\
gbianchini@frm.utn.edu.ar\\
ORCiD ID: 0000-0003-3609-9076
}
}

\maketitle

\begin{abstract}
  Wildfires are a highly prevalent multi-causal environmental phenomenon.
  The impact of this phenomenon includes human losses, environmental damage and high economic costs.
  To mitigate these effects, several computer simulation systems have been developed in order to predict fire behavior based on a set of input parameters, also called a scenario (wind speed and direction; temperature; etc.).
  However, the results of a simulation usually have a high degree of error due to the uncertainty in the values of some variables, because they are not known, or because their measurement may be imprecise, erroneous, or impossible to perform in real time.
  Previous works have proposed the combination of multiple results in order to reduce this uncertainty.
  State-of-the-art methods are based on parallel optimization strategies that use a fitness function to guide the search among all possible scenarios.
  Although these methods have shown improvements in the quality of predictions, they have some limitations related to the algorithms used for the selection of scenarios.
  To overcome these limitations, in this work we propose to apply the \emph{Novelty Search} paradigm, which replaces the objective function by a measure of the novelty of the solutions found, which allows the search to continuously generate solutions with behaviors that differ from one another.
  This approach avoids local optima and may be able to find useful solutions that would be difficult or impossible to find by other algorithms.
  As with existing methods, this proposal may also be adapted to other propagation models (floods, avalanches or landslides).
\end{abstract}

\begin{IEEEkeywords}
%
%
wildfire propagation prediction, evolutionary algorithms, novelty search, uncertainty reduction
\end{IEEEkeywords}

\section{Introduction}

Forest fires are rapidly spreading fires that affect vegetated lands such as forests, plains, grasslands, pastures, among others. 
In particular, wildfires are unplanned, unwanted, or uncontrolled fires \cite{ForestFire,Wildfires}.
Their causes can be natural, i.e., due to climatic factors, for example, drought, low humidity, heat waves; or anthropic, caused, in many cases, by campfires and used cigarettes that are not properly extinguished, open-air dumps, land abandonment, preparation of grazing areas with fire, among other causes.
The effects of climate change increase the risk of fires due to extreme droughts and high temperatures.
Among the negative impacts, it is possible to mention the loss of human lives, evacuations, damage to flora and fauna, atmospheric emissions, soil erosion, and large economic impacts, both in terms of damage repair and firefighting costs, as well as indirect damage to different productive activities (for example, the degradation of landscapes, which affects tourism).

The fight against this type of fires involves different phases, including detection of fire outbreaks, prediction of occurrence and prediction of behavior.
The methods referred to in this paper can be used in the prevention and prediction phases.
The tools for predicting the behavior of forest fires are of great interest for decision-making in fire control, in order to mitigate the damage.
There are several simulators developed to predict fire behavior, e.g., \emph{BEHAVE} \cite{burganBEHAVEFireBehavior1984}, \emph{FARSITE} \cite{finneyFARSITEFireArea1998}, \emph{fireLib} \cite{smithVFireLibForestFire2016}, \emph{BehavePlus} \cite{heinschBehavePlusFireModeling2010} and \emph{FireStation} \cite{lopesFireStationIntegratedSoftware2002}.
These simulators use propagation models whose objective is to predict the evolution of the fire line over a period of time.
This can be done either in real time (provided that the system meets the efficiency requirements), or as a preventive tool that can provide information like the detection of high risk areas or patterns of propagation with high probability.
For this, simulators require a set of input parameters, also called \emph{scenarios}, which describe environmental characteristics that affect propagation, such as the type of fuel material, its humidity, slope and relief of the terrain, temperature, and wind speed and direction.

From a computational point of view, this prediction problem is challenging due to the complexity of the models involved and the sources of uncertainty in the input data.
This last aspect is of great importance, since limitations in providing correct input parameters for the model lead to errors in the prediction.
This uncertainty is due to the difficulty or impossibility of obtaining the values of the variables, either because of resource limitations (e.g., number of sensors), because the measurements are indirect (e.g., vegetation moisture measurement), or because the variables have a dynamic behavior and their observation in real time is not feasible (e.g., wind characteristics).
This makes prediction based on a single solution unreliable.

For this reason, different methods have been proposed, which perform multiple simulations on different scenarios in order to detect trends and thus reduce such uncertainty.
One strategy, categorized as \emph{Data-Driven Methods} (DDMs), consists in the use of multiple simulations in order to select the set of parameters which obtains a better prediction of past fire behavior, and use it as input for the prediction of the following time step.
Examples of this strategy are found in \cite{abdalhaqClassicalIdealEnhancing2006,pinolModelCalibrationUncertainty2002}.
These methods are limited since they use only one scenario for uncertainty reduction, which may not yield good quality results due to the problems described above.
To overcome this, other approaches have started to combine results from multiple simulations in order to produce a prediction.
These are called \emph{Data-Driven Methods with Multiple Overlapping Solutions} (DDM-MOS). 
Given the high complexity of the space of scenarios, several DDM-MOS consider a reduced set, selected during an \emph{Optimization Stage}.
An example of such a method is called \emph{Evolutionary Statistical System} (ESS), which uses an evolutionary algorithm in order to find plausible scenarios for making a prediction.
Two recent proposals based on ESS are ESSIM-EA and ESSIM-DE. Both of them use different Parallel Evolutionary Algorithms (PEAs): a \emph{Genetic Algorithm} \cite{IntroductionGeneticAlgorithms,goldbergGeneticAlgorithmsSearch1988}, and \emph{Differential Evolution} \cite{bilalDifferentialEvolutionReview2020}, respectively. The objective of these algorithms is to reduce the space of scenarios to be considered, achieving improvements in predictive quality.
These approaches are guided by an objective function that, in other domains, is usually intended to converge to a single solution.
In problems where there is a high degree of uncertainty, this objective function is not always a direct indicator of the quality of the solutions.
The previous methods have encountered limitations such as premature convergence, and calibration and tuning techniques have been required in order to incorporate more diverse solutions into the prediction process.

Such results are indicators of the high complexity of the problem, which may be due to different factors.
For example, the fitness function could produce a search space with multiple local optima, leading the algorithm to get stuck and lose the possibility of reaching the global optimum or (at least) better local optima.
There are many other characteristics that may hinder the ability of an algorithm to produce optimal results \cite{malanSurveyTechniquesCharacterising2013a}.
Apart from these issues, there are other challenges related to particular domain, since the population-based metaheuristics used for this application have been modified in order to return a set of results instead of a single solution, which may imply that the solutions are similar to one another due to the convergence tendencies of these type of algoritms.
%
Based on this analysis, this work proposes a new method for the Optimization Stage that avoids the issues mentioned by the use of a different paradigm for guiding the search, i.e., the \emph{Novelty Search} (NS) paradigm \cite{lehmanAbandoningObjectivesEvolution2011,lehmanExploitingOpenEndednessSolve2008,lehmanEvolvabilityInevitableIncreasing2013}.
NS is an alternative approach that ignores the objective as a guide for exploration and instead rewards candidate solutions that exhibit novel (different from previously discovered) behaviors, in order to maximize exploration and avoid local optima.
Due to the characteristics of NS, we believe it is a promising alternative to the limitations of the metaheuristics previously applied to this problem.
The hypothesis of this work can be summarized as the idea that the application of a novelty-based metaheuristic to the fire propagation prediction problem can obtain comparable or better results in quality with respect to existing methods.
Although this new proposal is still in the implementation stage, we are optimistic as regards its theoretical guarantees.

In the next section we delve into previous works related to our present contribution; firstly, in the area of DDM-MOS, with a summary of previous systems (Sections \ref{sec:ess}~and~\ref{sec:essim}), and secondly, in the field of NS (Section \ref{sec:noveltyintro}), where we explain the paradigm and its related contributions in general terms.
Then, in Section~\ref{sec:novelty}, we present a detailed description of the proposed method, and provide a pseudocode of the algorithm.
Finally, in Section \ref{sec:conclusions}, we list our main conclusions and outline our future work.

\section{State-of-the-art}\label{sec:sota}


\subsection{General ESS system environment}\label{sec:ess}

\begin{figure*}[t!]
  \centering
  \includegraphics[scale=0.3]{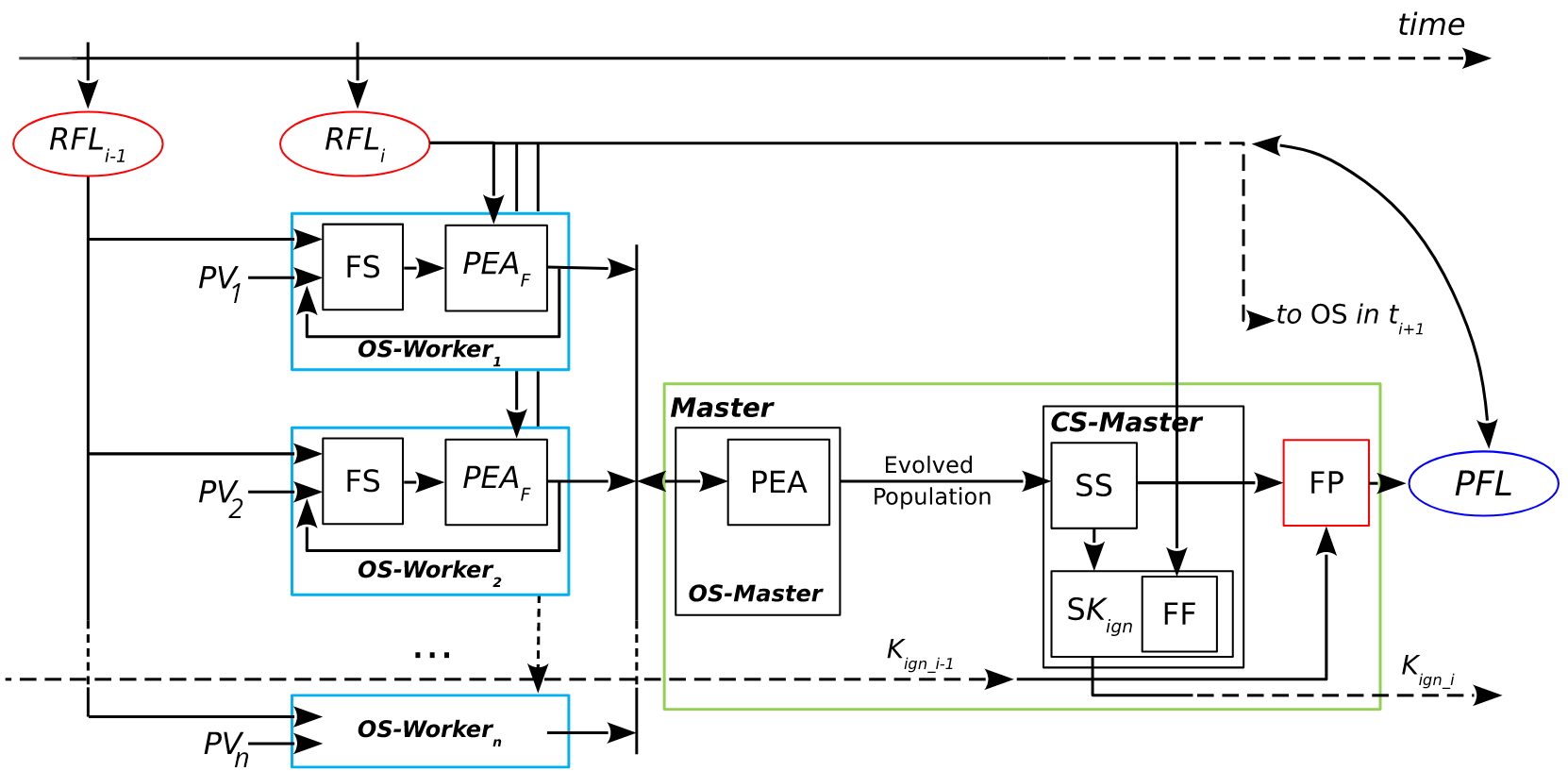}
  \caption{Evolutionary Statistical System. $FS$: fire simulator, \emph{OS-Master}: Optimization Stage in Master, \emph{OS-}$Worker_x$: Optimization Stage in Worker $x$, $PEA$: Parallel Evolutionary Algorithms,
$PEA_F$: Parallel Evolutionary Algorithm (fitness evaluation),
\emph{CS-Master}: Calibration Stage in Master,
$FF$: fitness function,
$RFL_i$: real fire line of instant $t_i$,
$PFL_i$: predicted fire line of instant $t_i$,
$PV_{\{1\dots n\}}$: parameter vectors (scenarios),
$t_n$: time instant $n$,
$K_{ign}$: \emph{Key Ignition Value} for $t_i$,
$FP$: Fire Prediction stage,
$SS$: Statistical Stage,
$SK_{ign}$: \emph{Key Ignition Value} search.}
\label{fig:ess}
\end{figure*}

To address the uncertainty present in the domain, several methods have been proposed with the aim of improving the quality of predictions.
The ones related to this work are classified as DDM-MOS, given that they perform multiple simulations, each based on a different scenario, and use a combination of simulations in order to obtain a prediction.
One such system, ESS, produces a combination of results based on an evolutionary algorithm that performs a search over the space of all possible scenarios, with the aim of reducing the complexity of the computations while considering a sample of scenarios that may yield better prediction results.
In order to understand the scheme of the method proposed in this paper, it is necessary to have a general comprehension of how ESS works, since they share a similar framework. 
A general scheme of ESS operation can be seen in Fig.~\ref{fig:ess} (reproduced with permission from \cite{mendezgarabettiComparativeAnalysisPerformance2015}).
In this system, a Master/Worker parallel design pattern is employed in order to reduce computation times.
The process considers different discrete time instants for fire propagation prediction.
These instants are called \emph{prediction steps} and, in each of them, four main stages take place: \underline{O}ptimization \underline{S}tage (divided into Master and Workers, respectively: \textbf{OS-Master} and \textbf{OS-Worker}), \underline{S}tatistical \underline{S}tage (\textbf{SS}), \underline{C}alibration \underline{S}tage (\textbf{CS-Master}) and \underline{P}rediction \underline{S}tage (\textbf{PS}).

The process starts with the \textbf{OS}, during which a scenario search is performed based on the fitness function.
The process starts with the initialization of the population, the evolution of the population (selection, reproduction and replacement) and the termination.
The Master process initializes and manages the population of scenarios, represented by a set of parameter vectors \textbf{PV}$_{\{1\dots n\}}$ (representing the scenarios or individuals), and then distributes these scenarios to the Workers processes.
The Workers use the scenario, along with the real fire line at instant $t_{i-1}$ ($LFR_{i-1}$), sending them as input to a fire behavior simulator \textbf{FS}, obtaining a simulated map for each scenario.
The fitness is obtained by comparing the simulated map (produced by the simulator based on a scenario) with the real state of the fire at that instant of time.
After obtaining the simulation, the simulated map is compared with the real state of the fire at the same instant of time, $LFR_i$, according to the fitness function.

\begin{figure*}[t!]
  \centering
  \includegraphics[scale=0.56]{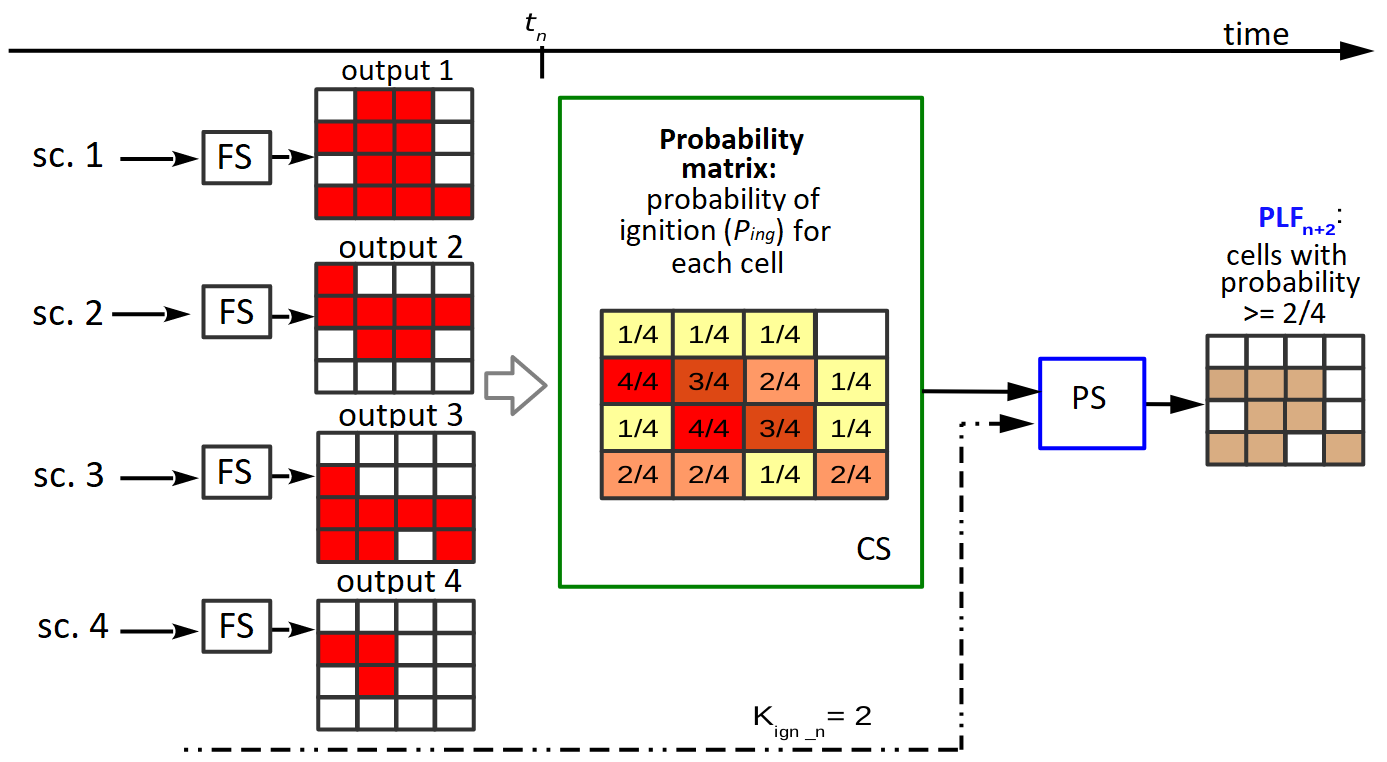}
  \caption{Generation of the prediction. \emph{PS}: Prediction Stage. \emph{CS}: Calibration Stage. \emph{sc.}: scenario, $FS$: fire simulator, $K_{ign\_n}$: \emph{Key Ignition Value} computed for instant $t_n$. $PFL_t$: predicted fire line for $t$.}
\label{fig:matrix}
\end{figure*}

Once the \textbf{OS-Master} has finished, the following stages are \textbf{CS} and \textbf{PS}.
These stages are further illustrated in Fig.~\ref{fig:matrix} (reproduced and translated with permission from \cite{tardivoParalelizacionSintonizacionEvolucion}). 
The first step is for the Master to aggregate the resulting maps into a matrix in which each cell represents the probability of ignition of that region.
This corresponds to the Statistical Stage or \textbf{SS}.
Such a matrix will be used for two purposes: on the one hand, it is provided as input for the Prediction Stage; on the other hand, it is used in the Calibration Stage (\textbf{CS-Master}).
The \textbf{CS-Master} is needed to obtain a prediction from the aggregated map.
At this stage, a probability map is computed to obtain a threshold value called \emph{Key Ignition Value}, or $K_{ign}$, which best represents the fire behavior pattern for the given simulation step.
This value is obtained by searching for a threshold value that, when applied to the probability matrix, produces the best prediction in terms of the fitness function for the current time step.
This search is represented by the $SK_{ign}$ block in Fig.~\ref{fig:ess}.
The new value $K_{ign}$ is used within the \textbf{PS} of the next prediction step; therefore, the prediction cannot start at the first time instant.
Once a first threshold value was obtained, it is possible to perform the \textbf{PS} at each subsequent iteration, for which the matrix obtained by applying the threshold $K_{ign_n}$ is used to perform the fire line prediction for the current time step.
The new value $K_{ign_n+1}$ will be used in the next prediction step.

\subsection{Operation of ESSIM-EA and ESSIM-DE}\label{sec:essim}


Both ESSIM-EA and ESSIM-DE are based on ESS; therefore, in general terms, the predictive process is divided into the same stages.
However, since both systems use a two-level hierarchical process scheme, these stages are subdivided differently according to this hierarchy: Monitor, Masters and Workers.
Essentially, the system uses a number of islands, each of which has a Master and a number of Workers; the Monitor acts as the Master process for the Masters of the islands.

At the beginning of the predictive process, the Monitor sends to each island the initial information to carry out the different stages.
The Master process of each island performs the \textbf{OS}, managing the evolutionary process, migration, and completion (return of results to the Monitor).
On each island, the Master sends individuals to the Workers processes, which are in charge of their evaluation.
After the evaluations, the Master performs the \textbf{SS}, obtaining the probability matrix for the \textbf{CS} and \textbf{PS}.
The \textbf{SS} and \textbf{CS} are carried out by the Master; while the \textbf{PS} is performed by the Monitor.
This process receives all the probability matrices generated by the Masters, together with their $K_{ign}$ value and the associated \emph{fitness}, calculated by \eqref{eq:fitness}.
The Monitor then selects the best candidate, producing the current step prediction.

Experimentally, ESSIM-EA has been shown to obtain good quality predictions, while ESSIM-DE significantly reduced response times, but did not obtain quality improvements.
For this reason, there have been subsequent works aiming to improve the performance of the ESSIM-DE method by means of \emph{tuning strategies} \cite{naonoSoftwareAutomaticTuning2010}.
Tuning strategies allow for the calibration some critical aspect, bottleneck or limiting factor of an application to improve its performance.
These can be automatic (when the techniques are transparently incorporated in the application) and/or dynamic (adjustments occur during execution) \cite{caymesscutariEnvironmentAutomaticDevelopment2016}.
In the case of ESSIM-DE, two automatic and dynamic tuning metrics were developed and implemented, both aimed at mitigating the issues of premature convergence and population stagnation present in the case of application of the algorithm.
One metric was a \emph{population restart operator} \cite{tardivoOptimizationUncertaintyReduction2018}, and the other involved the analysis of the IQR factor of the population throughout generations \cite{caymesscutariDynamicTuningForest2020}.
The results showed that ESSIM-DE enhanced with these metrics achieved better quality and response times with respect to the same method without tuning.

Despite the improvements demonstrated by both approaches, they still have some limitations.
Firstly, the design of the Optimization Stage in ESSIM-EA is based on metaheuristics intended for single solution problems, which use a fitness function to evaluate the quality of the solutions.
In this system, the solutions of the last generated population are used to select the set of solutions to be used in the prediction stages.
Evolutionary metaheuristics tend to converge to a population of similar genotypes, that is, of individuals which are similar in their representation, which in this case is the set of values of the parameters.
Thus, the population evolved for each prediction step may consist of a set of scenarios similar to each other, which limits the contribution of these solutions to uncertainty reduction and defeats its purpose.
In complex problems, it is often the case that different solutions may be genotypically far apart in the search space, but may still have acceptable fitness values that contribute to the prediction.
Therefore, algorithms intended for a single solution may leave out these promising candidates.
Secondly, in the case of ESSIM-DE, in a first version it was found that the quality of the results did not improve with respect to ESSIM-EA, so it was modified to a new version that tends toward greater diversity, where a part of the results are incorporated in the prediction process regardless of their fitness.
This modification produced better results in quality than the original version, but the issues of stagnation and premature convergence remained. 
This led to the need to design the aforementioned tuning mechanisms, which may compensate for the limitations of the chosen algorithm, but they may also discard potentially useful information from populations produced in earlier prediction steps.

\subsection{The Novelty Search paradigm}\label{sec:noveltyintro}

Traditional search approaches, including metaheuristics such as those used in ESS, ESSIM-EA and ESSIM-DE, reward the degree of progress toward a solution, measured by an objective function, usually referred to as a \emph{score} or \emph{fitness function}, depending on the type of algorithm.
This function, together with the neighborhood operators of a particular algorithm, produces what is known as the \emph{fitness landscape}.
In highly complex problems, the \emph{fitness landscape} often has features that make the objective function inefficient in guiding the search, or even prevent the search from finding good solutions.
A particularly difficult example of the latter is a characteristic known as \emph{deceptiveness}.
In simple terms, an objective function is deceptive with respect to a given algorithm when the combination (through the operators of the algorithm) of solutions of high fitness leads to solutions of lower fitness and vice versa \cite{ollionWhyHowMeasure2011a,malanSurveyTechniquesCharacterising2013a}.
These limitations in traditional search approaches led to the creation of alternative strategies that address the limitations inherent to objective-based methods.
One of these strategies is NS, introduced in \cite{lehmanExploitingOpenEndednessSolve2008}. 
In this paradigm, the search is driven by a characterization of the behavior of individuals that rewards the dissimilarity of new solutions with respect to those previously explored.
As a consequence, the search never converges but rather explores many different regions of the search space, which allows the search to discover high fitness solutions that may not be reachable by traditional algorithms, including metaheuristics such as fitness-based evolutionary algorithms.
NS has been applied with good results to multiple problems from diverse fields \cite{gomesEvolutionSwarmRobotics2013,krcahSolvingDeceptiveTasks2012,lehmanExploitingOpenEndednessSolve2008,lehmanAbandoningObjectivesEvolution2011,lehmanEvolvingDiversityVirtual2011,ollionWhyHowMeasure2011}.
It is interesting to note that there are two main areas for the application of this paradigm: firstly, in open-ended problems, whose objective is the generation of increasingly complex, novel or creative behaviors, without a unique or predetermined solution to be reached; secondly, it also serves for optimization problems in general, allowing to find global optima in many cases, and outperforming traditional metaheuristics when the problems are deceptive.

In order to measure the novelty of solutions, algorithms following this paradigm need to implement a function to evaluate the \emph{novelty score} of the solutions.
For this, a distance measure $dist$ must be defined in the space of behaviors of the solutions.
This measure is problem-dependent; an example can be the difference between values of the fitness function of two individuals.
A frequently used novelty score function is the one presented by \cite{lehmanExploitingOpenEndednessSolve2008}. For an individual $x$, it computes the average distance to its $k$ closest neighbors: 

\begin{equation}\label{eq:noveltyscore}
  \rho(x)=\frac{1}{k} \sum_{i=0}^{k-1} {dist(x,\mu_i)},
\end{equation}

where $\mu_i$ is the $i$-th nearest neighbor of the individual $x$ according to the distance measure $dist$.
In the literature, the parameter $k$ is usually selected experimentally, but the entire population can also be used \cite{lehmanAbandoningObjectivesEvolution2011,gomesDevisingEffectiveNovelty2015}.

To perform this evaluation, it is not sufficient to select close individuals by considering only the current population; it is also necessary to consider the set of individuals that have been novel in past iterations.
To this end, the search incorporates an \emph{archive of novel solutions} that allows it to keep track of the most novel solutions discovered so far, and uses it to compute the novelty score.
The novelty values obtained are used (instead of the traditional fitness-based score) to guide the search in a way that maximizes exploration of the search space.
This design allows the search to be unaffected by the fitness function landscape, directly preventing problems such as those found in the systems described in the previous section.
When using conventional metaheuristics, due to the randomness involved in the algorithms, it is possible that some high fitness solutions may be lost in intermediate iterations, with no record of them remaining in the final population.
In contrast, NS can avoid this issue because, when applied to optimization problems, it makes use of a memory of the best performing solution(s), as measured, for example, by the fitness function.
In this way, even though NS never converges to populations of high fitness, it is possible to keep track of the best solutions (with respect to the fitness function or any characterization of the behavior of the solutions) found throughout the search.

The limitations observed in existing systems led us to consider the selection of other search approaches that may yield improvements in the quality of predictions.
Given the particular problems observed in the experimental results of previous works, a promising approach for this problem could be NS.
Different metaheuristics have already been implemented using the NS paradigm, such as a \emph{Genetic Algorithm} \cite{doncieuxNoveltySearchMakes2020} and \emph{Particle Swarm Optimization} \cite{galvaoNoveltyDrivenParticleSwarm2015}.
Additionally, multiple hybrid approaches that combine fitness and novelty exist in the literature and have been shown to be effective in solving practical problems.
Among some of the approaches used, there are weighted sums between fitness and novelty-based goals \cite{cuccuWhenNoveltyNot2011}, different goals in a multi-objective search \cite{mouretEncouragingBehavioralDiversity2012}, independent searches with some type of interaction among each other \cite{krcahSolvingDeceptiveTasks2012}, among others \cite{pughConfrontingChallengeQuality2015, cullyRobotsThatCan2015,mouretIlluminatingSearchSpaces2015,lehmanExploitingOpenEndednessSolve2008,lehmanEvolvingDiversityVirtual2011,gomesEvolutionSwarmRobotics2013}.

Another important aspect of this type of approach is that, similarly to the metaheuristics of current prediction systems, novelty-based search algorithms also allow for various parallelization possibilities.
An example of a parallelization approach using NS can be found in \cite{hodjatDistributedAgeLayeredNovelty2016}.
In general, metaheuristics can usually be parallelized by means of one or more parallelization models in order to obtain better execution times, more efficient use of resources, or improvements in the performance of the algorithm.
In the particular case of NS, parallelization can be a way to allow the search a large exploration capacity without excessively affecting the execution time.
Furthermore, since it can be implemented by adapting traditional metaheuristics, NS can also be subject to the same parallelization models as these.
However, by differing from traditional methods by the inclusion of an archive and by the possibility of hybridization with fitness-based techniques, new opportunities and challenges arise in the design of parallel methods.

\section{Novelty-based approach for uncertainty reduction in wildfire prediction}\label{sec:novelty}

\subsection{Integration of Novelty Search into the prediction framework}\label{sec:noveltyframework}

\begin{figure*}[t!]
  \centering
  \includegraphics[scale=0.28]{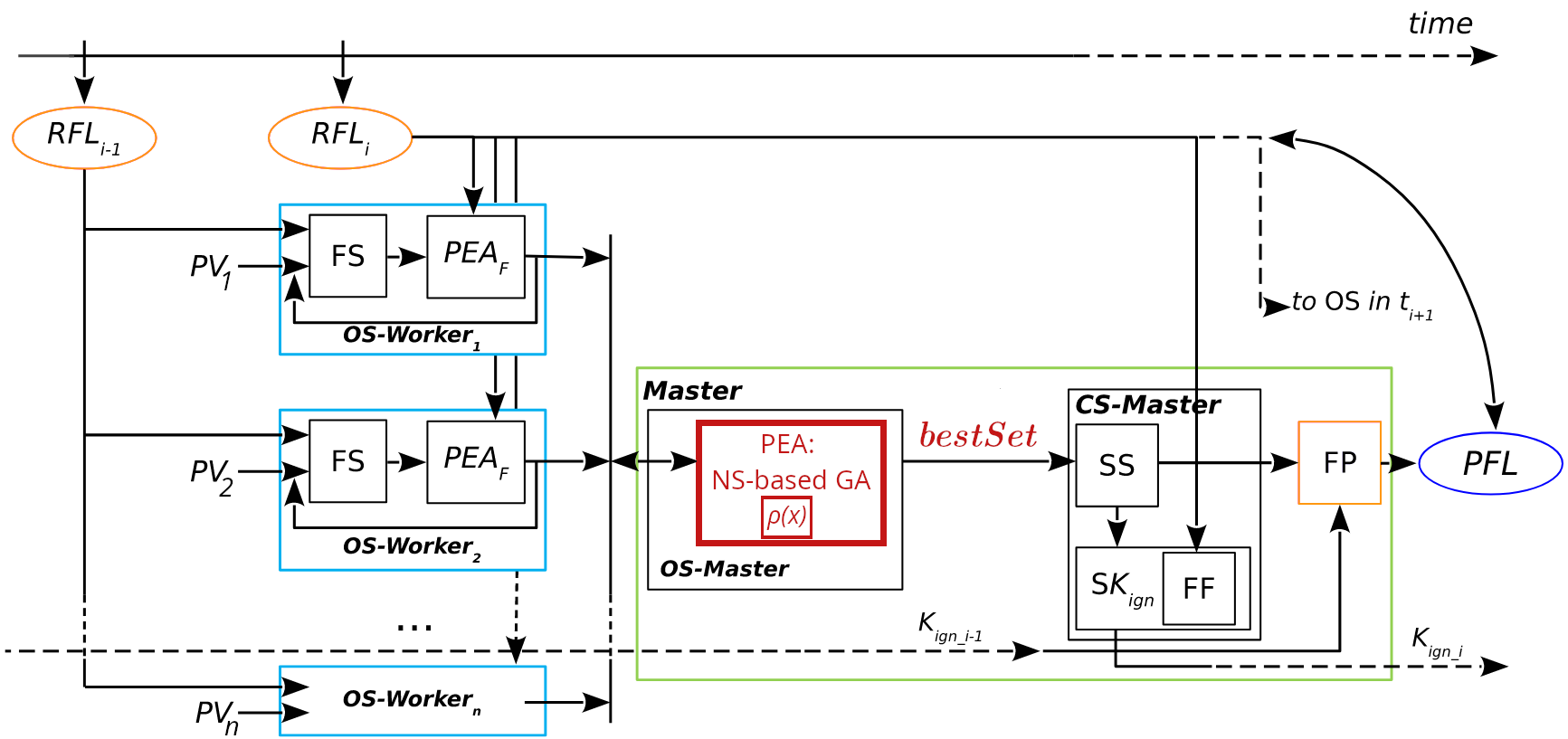}
  \caption{Evolutionary Statistical System - Novelty Search. $FS$: fire simulator, \emph{OS-Master}: Optimization Stage in Master, \emph{OS-}$Worker_x$: Optimization Stage in Worker $x$, $PEA$: Parallel Evolutionary Algorithm, \emph{NS-based GA}: Novelty Search-based Genetic Algorithm, $\rho(x)$: novelty score function \eqref{eq:noveltyscore}, $PEA_F$: Parallel Evolutionary Algorithm (fitness evaluation), \emph{CS-Master}: Calibration Stage in Master, $FF$: fitness function, $RFL_i$: real fire line of instant $t_i$, $PFL_i$: predicted fire line of instant $t_i$, $PV_{\{1\dots n\}}$: parameter vectors (scenarios), $t_n$: time instant $n$, $K_{ign}$: \emph{Key Ignition Value} for $t_i$, $FP$: Fire Prediction stage, $SS$: Statistical Stage,
$SK_{ign}$: \emph{Key Ignition Value} search.
}
\label{fig:essns}
\end{figure*}

Our preliminary framework for the new method is illustrated by Fig.~\ref{fig:essns}.
It presents some aspects that are unchanged with respect to ESS (compare to Fig.~\ref{fig:ess}), while others have important modifications.
Therefore, we have highlighted these relevant changes in the figure. 
This framework, called ESS-NS (\emph{Evolutionary Statistical System - Novelty Search}), consists of a process that follows the same stages presented in Fig.~\ref{fig:ess} (Section~\ref{sec:ess}): \underline{O}ptimization \underline{S}tage (\textbf{OS}), \underline{S}tatistical \underline{S}tage (\textbf{SS}), \underline{C}alibration \underline{S}tage (\textbf{CS}), and \underline{P}rediction \underline{S}tage (\textbf{PS}).
It also uses the same propagation simulator, called \emph{fireLib} \cite{smithVFireLibForestFire2016}, which is implemented in an open source and portable library.
The input parameters for the simulator are: a map of the terrain, and the set of parameters concerning the environmental conditions and terrain topography. 
A description of these parameters and their characteristics is shown in Table~\ref{tab:parameters}. In the first row, the \emph{Rothermel Fuel Model} refers to a taxonomy that characterizes 13 models of fire propagation, which is commonly used by a number of simulators, including \emph{fireLib}. For more information on the parameters considered in this library, see \cite{andrewsBehavePlusFireModeling2009}.
\begin{table}
  \footnotesize
  \centering
  \begin{tabular}{|l|p{3.3cm}|l|p{1.5cm}|}
    \hline
Parameter & Description & Range & Unit of measurement \\ \hline
Model & Rothermel Fuel Model & 1-13 & fuel model \\
WindSpd & Wind speed & 0-80 & miles/hour \\
WindDir & Wind direction & 0-360 & degrees clockwise from North \\
M1 & Dead Fuel Moisture in 1 hour since start of fire & 1-60 & percent \\
M10 & Dead Fuel Moisture in 10 h & 1-60 & percent \\
M100 & Dead Fuel Moisture in 100 h & 1-60 & percent \\
Mherb & Live herbaceous fuel moisture & 30-300 & percent \\
Slope & Surface slope & 0-81 & degrees \\
Aspect & Direction of the surface faces & 0-360 & degrees clockwise from north \\\hline
  \end{tabular}
  \vspace*{1mm}
  \caption{Parameters used by the \emph{fireLib} library.}
  \label{tab:parameters}
\end{table}

The output is another map indicating the time instant of ignition of each cell, that is, the moment when that cell is reached by the fire, or zero otherwise.
To a great extent, the order and functioning of the stages and their assignment to Master and Workers are maintained.

Regarding the aspects that are modified, there are two general differences in the parallel framework.
First, it replaces the metaheuristic in the \textbf{OS}, by which the search method is still an evolutionary algorithm, but one that implements a novelty-based strategy using a genetic algorithm, as shown in the red block (\emph{PEA: NS-based GA}) in Fig.~\ref{fig:essns}.

This algorithm is described in detail in the following section, but it is important to note here that there is an additional computation of a score, that is, the \emph{novelty score}, represented by the function $\rho(x)$ from \eqref{eq:noveltyscore}.
Second, the output of the optimization algorithm is not the final evolved population, as in previous methods; rather, it is a collection of high fitness individuals which were accumulated during the search, which we call $bestSet$.
This has to do with the fact that NS does not converge, and it is its main advantage for this application: it has the ability to record individuals from completely different areas of the search space, and include them for the construction of the aggregated matrix.
In this way, we can introduce more valuable scenarios that may be very different among each other, and this might help reduce the uncertainty in the prediction process.

In addition, when compared to the more recent systems, ESSIM-EA and ESSIM-DE, the two-level hierarchical model is simplified back to a one-level Master/Worker model (with no islands) in which the Master process only delegates the simulation and evaluation of individuals to the Workers, since this is the most demanding part of the prediction process.
Because of this, the resulting framework is very similar to that of ESS.
This simplification is motivated by the need to have a baseline algorithm for future comparisons, and to be able to analyse the impact of NS alone on the quality of results.
Regarding this aspect, it is important to note that, in and of itself, NS uses a strategy that not only keeps diversity but actively reinforces it, and therefore, it would solve the problem that originally made it necessary to resort to mechanisms such as the island model.
Also, a more complex design would require additional decisions regarding the behavior of the algorithm with respect to the representation of the population and the fitness function, and these decisiones can directly affect both the quality and the efficiency of the method.
For example, a design such as an island model would require the specification of several criteria in order to perform migrations, e.g., a selection mechanism for individuals to emigrate and another mechanism to incorporate new individuals into the island of destination.
At the moment, these considerations are left as future work.

\subsection{Algorithm design and discussion}\label{sec:noveltyapplication}

The present proposal consists of applying an evolutionary metaheuristic based on novelty search as part of the Optimization Stage of a wildfire prediction system.
We have selected as metaheuristic a classical genetic algorithm (GA), adapted to the NS paradigm, both for simplicity of implementation and for comparative purposes with existing systems, which are both based on variants of evolutionary algorithms.

The novelty measure selected is computed as in~\eqref{eq:noveltyscore}. In this context, $x$ is a scenario, and we define $dist$ as the difference between the fitness values of each pair of scenarios:

\begin{equation}
  dist(x,\mu_i) = fitness(x) - fitness(\mu_i),
  \label{eq:distfitness}
\end{equation}

The fitness function used for this purpose is the one used in the ESS system and its successors: the Jaccard Index \cite{realProbabilisticBasisJaccard1996}, by considering the map of the field as a matrix of square cells (which is the representation used by the simulator):

\begin{equation}
  fitness(A,B) = \frac{A\cap B}{A \cup B},
  \label{eq:fitness}
\end{equation}

where $A$ represents the set of cells in the real map without the subset of burned cells before starting the simulations, and $B$ represents the set of cells in the simulation map without the subset of burned cells before starting the simulation.
Previously burned cells are not considered in order to avoid skewed results.
This formula measures the similarity between prediction and reality, and is equal to one when there is a perfect prediction, while a value of zero indicates the worst prediction possible.

The pseudocode of the proposed solution is found in Algorithm~\ref{a:nsgams}.
The high-level procedure is partially inspired on the algorithm proposed in \cite{doncieuxNoveltySearchMakes2020}, differing in several aspects.
The main difference is that our version adds a collection of solutions, $bestSet$.
This collection is updated at each iteration of the GA, so that at the conclusion of the main loop of the algorithm, the resulting set contains the solutions of highest fitness found during the entire search.
It should be noted that this set is used as resulting set, instead of the evolved population set which is used by the previous evolutionary-based systems, for both the \textbf{CS} and \textbf{PS}.
In addition, this algorithm uses two stopping conditions (line \ref{a:while}): by number of generations and by a threshold of fitness (both present in ESSIM-EA and ESSIM-DE), and also specifies conventional GA parameters, such as mutation rate and crossover.
These parameters are specified as input to the algorithm.
Another difference is that the archive of novel solutions ($archive$) is managed with replacement based on novelty only, as opposed to the pseudocode in \cite{doncieuxNoveltySearchMakes2020}, which uses a randomized approach.  
These features correspond to a ``classical'' implementation of the NS paradigm: an optimization guided exclusively by the novelty criterion, and a set of results based on the best values obtained using the fitness function.
These criteria allow us to establish a baseline against which it will be possible to perform comparisons among future variants of the algorithm.
In a first version, parallelism will only be implemented in the evaluation of the scenarios, i.e., in the simulation process and subsequent computation of the fitness function.


We now provide a detailed description of Algorithm~\ref{a:nsgams}.
\begin{algorithm}[t!]
  \caption{\label{a:nsgams}Novelty-based Genetic Algorithm with Multiple Solutions.}
  {\footnotesize
  \begin{algorithmic}[1]
  \REQUIRE population size $N$, number of offspring $m$, mutation rate $mR$, crossover rate $cR$, number of neighbors for novelty score $k$, maximum number of generations $maxGen$, fitness threshold $fThreshold$
    \ENSURE the set $bestSet$ of individuals of highest fitness found during the search
      \STATE $population \longleftarrow initializePopulation(N)$
    \STATE $archive \longleftarrow \emptyset$\label{a:archiveinit}
    \STATE $bestSet \longleftarrow \emptyset$\label{a:bestsetinit}
    \STATE $generations \longleftarrow 0$\label{a:geninit}
    \STATE $\maxFitness \longleftarrow 0$\label{a:maxfitinit}
    \WHILE {$generations < maxGen$ \AND $\maxFitness < fThreshold$}\label{a:while}%
    \STATE $\offspring \longleftarrow \generateOffspring(population, m, mR, cR)$ \label{a:genoffspring}
      \FOR {each individual $ind \in (population \cup \offspring$)}\label{a:forfitness}%
        \STATE $\indfitness \longleftarrow \evaluateFitness(ind)$ \label{a:evalfitness}
      \ENDFOR \label{a:endforfitness}
      \STATE $noveltySet \longleftarrow (population \cup \offspring\ \cup\ archive)$ \label{a:noveltyset}
      \FOR {each individual $ind \in (population \cup \offspring)$}\label{a:fornovelty}%
        \STATE $ind.novelty \longleftarrow evaluateNovelty(ind, noveltySet, k)$ \label{a:evalnovelty}
      \ENDFOR \label{a:endfornovelty}
      \STATE $archive \longleftarrow updateArchive(archive, \offspring)$ \label{a:updatearchive} 
      \STATE $population \longleftarrow replaceByNovelty(population, \offspring, N)$\label{a:replacepop}
      \STATE $bestSet \longleftarrow updateBest(bestSet, \offspring)$ \label{a:updatebest}
      \STATE $\maxFitness \longleftarrow \getMaxFitness(bestSet)$ \label{a:updatemaxfit}
      \STATE $generations \longleftarrow generations + 1$ \label{a:increasegen}
    \ENDWHILE \label{a:endwhile}
    \RETURN $bestSet$

  \end{algorithmic}
  }
\end{algorithm}
The input parameters of the algorithm include: the typical GA parameters, the two stopping conditions (maximum number of generations and fitness threshold), and one NS parameter: the number of neighbors to be considered for the computation of the novelty score in \eqref{eq:noveltyscore}.
The GA population selection strategy will be by roulette wheel selection, while the replacement strategy will be an elitist selection based on the whole population (the offspring will replace the parents depending on their novelty).
For the first version, we are considering a fixed size archive and solution set, but these sizes can be parameterized or even designed to dynamically change size during the search.
The output, $bestSet$, consists of a collection of solutions obtained throughout the search.
Each iteration of the main loop (lines \ref{a:while} to \ref{a:endwhile}) corresponds to a generation of the GA.
At the beginning of each generation, the algorithm performs the selection and reproduction steps, abstracted in $\generateOffspring$; that is, it generates $m$ offspring based on the current population.
The fitness computation, which is performed by the workers, is represented by the lines \ref{a:forfitness}~to~\ref{a:endforfitness}; it is calculated for all individuals.
These values are needed both for recording in $bestSet$ the best solutions and for the computation of each individual's novelty score from \eqref{eq:noveltyscore}.
Therefore, a second loop is needed for the novelty computation (lines \ref{a:fornovelty} to \ref{a:endfornovelty}).
Internally, $evaluateNovelty$ compares the individual $ind$ with each of the individuals in the reference set $noveltySet$ using the distance measure $dist$, and then takes the $k$ nearest neighbors, i.e., those individuals $ind'\in noveltySet$ for which the smallest values of $dist(ind,ind')$ are obtained, and uses them to evaluate the novelty function according to \eqref{eq:noveltyscore}, where $dist$ is computed by~\eqref{eq:distfitness}.
After this loop, the next two lines are the ones that define the search to be driven by the novelty score.
In the line \ref{a:updatearchive}, the archive of novel solutions is updated with the descendants that have higher novelty values. 
Population replacement is performed in line \ref{a:replacepop}, also using the novelty criterion.
Then, line \ref{a:updatebest} updates $bestSet$ to incorporate the solutions in $\offspring$ that have obtained better fitness values.
The lines \ref{a:updatemaxfit} and \ref{a:increasegen} update the maximum value of fitness found and the evolutionary generation number, respectively, which allow the algorithm to verify the stopping conditions in the next iteration.

\section{Conclusions and Future Work}\label{sec:conclusions}

We have proposed a new parallel metaheuristic approach for uncertainty reduction in the problem of wildfire propagation prediction.
This method might also be adapted for the prediction of other propagation phenomena such as floods, avalanches or landslides.
As next steps, we are working on the implementation and experimentation of the first version of this strategy.
This version uses parallelization only at the level of the fitness evaluations (which involves all simulations of scenarios).
It also focuses on remembering the solutions of highest fitness found during the search, a strategy that is expected to obtain better results if the fitness function is representative of the quality of solutions, regardless of how scattered these solutions may be throughout the search space.
There is also the possibility that high fitness solutions are not enough, due to limitations in the fitness function.
For example, rapidly changing conditions may entail that a scenario that was a good descriptor at one time step can become worse at the next step.
If such were the case, we may also explore possible variants of the algorithm that build a solution set not only according to fitness values but also by some criterion, like the addition of a percentage of novel or random solutions.
Other lines of future work are the implementation of parallel and/or distributed methods such as an island model, which may incorporate hybridization with fitness-based strategies.
There is also the possibility of using a dynamic size archive and/or solution set, a novelty threshold for including solutions in the archive as in \cite{lehmanExploitingOpenEndednessSolve2008}, or even switching the underlying metaheuristic and adapting its mechanisms to the application problem.


%



\bibliography{pdco2022}
\bibliographystyle{ieeetr}

\end{document}